\documentclass[a4paper]{jpconf}
\usepackage{graphicx}
\usepackage{lineno}
\begin{document}
\title{Topological studies of light-flavor hadron production in pp, p--Pb, and Pb--Pb collisions with ALICE at the LHC}

\author{Sushanta Tripathy (for the ALICE collaboration)}

\address{Instituto de Ciencias Nucleares, Universidad Nacional Aut\'onoma de M\'exico,\\
  Apartado Postal 70-543, M\'exico Distrito Federal 04510, M\'exico}

\ead{sushanta.tripathy@cern.ch}

\begin{abstract}
Recent results for high multiplicity pp and p-Pb collisions have revealed that they exhibit heavy-ion-like behaviors. To understand the origin(s) of these unexpected phenomena, event shape observables such as transverse spherocity ($S_{\rm 0}^{p_{\rm T} = 1}$) and the relative transverse activity classifier ($R_{\rm{T}}$) can be exploited as a powerful tools to disentangle soft (non-perturbative) and hard (perturbative) particle production. Here, the production of light-flavor hadrons is shown for various $S_{\rm 0}^{p_{\rm T} = 1}$ classes in pp collisions at $\sqrt{s}$ = 13 $\textrm{TeV}$ measured with the ALICE detector at the LHC are presented. The evolution of average transverse momentum ($\langle p_{\rm T}\rangle$) with charged-particle multiplicity, and identified particle ratios as a function of $p_{\rm T}$ for different $S_{\rm 0}^{p_{\rm T} = 1}$ are also presented. In addition, the system size dependence of charged-particle production in pp, p--Pb, and Pb--Pb collisions at $\sqrt{s_{\rm NN}}$ = 5.02 TeV is presented. The evolution of $\langle p_{\rm T}\rangle$ in different topological regions as a function of $R_{\rm{T}}$ are presented. Finally, using the same approach, we present a search for jet quenching behavior in small collision systems.
\end{abstract}

\section{Introduction}
Recent measurements by ALICE~\cite{ALICE:2017jyt} show a smooth evolution of strange to non-strange particle ratios across colliding systems (pp, p--Pb, and Pb--Pb) as a function of charged-particle multiplicities. The observed enhancement scales with the strangeness content of the hadron. Similar features of various particle ratios measured across different collision systems are also observed in Ref.~\cite{Acharya:2018orn}. Collective-like effects have also been observed in small systems (pp and p-Pb collisions), but there is so far no evidence for jet quenching in small systems~\cite{Nagle:2018nvi}. In popular event generators such as PYTHIA8~\cite{Sjostrand:2014zea} the observed effects are only reproduced when novel final-state interactions (ropes and shoving) are introduced. In EPOS-LHC~\cite{Pierog:2013ria}, one can qualitatively describe the effect as by inclusion of QGP production in small systems. Small systems in EPOS-LHC are a mix of corona (QCD) and core (QGP), and it is the relative fraction of these that mainly drives the change with system size. To understand the origins of these phenomena, event shape observables such as the unweighted transverse spherocity ($S_{\rm 0}^{p_{\rm T = 1}}$) and the relative transverse activity classifier ($R_{\rm{T}}$) can be exploited as a powerful tools to disentangle soft and hard particle production.

In this contribution, we report the production of light-flavor hadrons for various $S_{\rm 0}^{p_{\rm T} = 1}$ classes in pp collisions at $\sqrt{s}$ = 13 $\textrm{TeV}$. In addition, the system size dependence of charged particle production in pp, p--Pb, and Pb--Pb collisions at  $\sqrt{s_{\rm NN}}$ = 5.02 TeV as a function of $R_{\rm{T}}$ is reported. The evolution of $\langle p_{\rm T}\rangle$ in different topological regions as a function of $R_{\rm{T}}$ are presented. Finally, we present a search for jet quenching behavior in small collision systems using similar techniques.

\section{Event shape observables}
Event shape observables have the capability to separate events with back-to-back jet structures and events dominated by multiple soft scatterings. Here, we use $S_{\rm 0}^{p_{\rm T} = 1}$ and $R_{\rm T}$ as estimators to disentangle events with these topological limits.

\subsection{Unweighted transverse spherocity ($S_{\rm 0}^{p_{\rm T} = 1}$)}

The unweighted transverse spherocity, inspired from transverse spherocity~\cite{Acharya:2019mzb}, is given by,
\begin{eqnarray}
S_{0}^{p_{\rm T} = 1} = \frac{\pi^{2}}{4} \min_{\hat{n}} \bigg(\frac{\Sigma_{i}~|p_{\rm T_{i}}\times\hat{n}|}{N_{\rm trks}}\bigg)^{2},
\label{eq1}
\end{eqnarray}
where $N_{\rm trks}$ is total number of charged-particle tracks in a given event and $\hat{n}$ is a unit vector that minimizes Eq.~\ref{eq1}. Here, $S_{0}^{p_{\rm T} = 1}$ is calculated using charged-particle tracks that have $p_{\rm T} > $ 0.15 GeV/$c$ for events with at least 10 charged-particle tracks to ensure that the concept of a topology is statistically meaningful. The charged-particle tracks are reconstructed using the Time Projection Chamber (TPC), within the pseudorapidity interval $|\eta|<0.8$. Unlike the estimator discussed in Ref.~\cite{Acharya:2019mzb}, the $p_{\rm T}$ of each track is normalized to unity ($p_{\rm T}$ = 1). This modification in the estimator is required to minimize biases which affect neutral particle yields. By definition, the $S_{0}^{p_{\rm T = 1}}$ estimator varies between the values 0 and 1, where the two extreme limits correspond to the two different topological limits. Events with $S_{0}^{p_{\rm T} = 1} \rightarrow$ 0 are dominated by a single back-to-back jet-like structure while events with $S_{0}^{p_{\rm T} = 1} \rightarrow$ 1 are dominated by isotropic particle production. From here onwards, the events within the bottom 20\% quantile of the $S_{0}^{p_{\rm T} = 1}$ distribution are referred to as the jetty events while the top 20\% quantile of the $S_{0}^{p_{\rm T} = 1}$ distribution are referred to as the isotropic events. As reported by ALICE~\cite{ALICE:2017jyt}, the strangeness enhancement is observed in high-multiplicity pp collisions, in addition to the $S_{0}^{p_{\rm T} = 1}$ selection, a top 10\% high-multiplicity requirement is also imposed for the event selection (multiplicity class I-III~\cite{Acharya:2019mzb}). The multiplicity is estimated using two sub-detector systems in ALICE. The multiplicity is estimated at mid-rapidity ($|\eta|<0.8$) using the Inner Tracking System (ITS) and it is referred as $N_{\rm SPD}$ estimator. The multiplicity is also estimated using the V0M scintillators which measure the charged-particle multiplicity at forward rapidity (2.8 $<\eta<$ 5.1 and -3.7 $<\eta<$ -1.7) and  it is referred as V0M estimator. The major difference in the measurements of particle production using two different estimators is that for V0M-based selections, the pseudo-rapidity regions for multiplicity estimation and for the measurement of the particle spectra are different, while for $N_{\rm SPD}$-based selections, the multiplicity and particle spectra are measured in the same rapidity region. 

\subsection{Relative transverse activity classifier ($R_{\rm T}$)}
The measurements of particle production as a function of relative transverse activity classifier allow for the estimation of the production of hard probes relative to the underlying event. To ensure that a hard scattering took place in the event, analysed events are required to have a leading trigger particle of 8 $< p_{\rm T} <$ 15 GeV/$c$. An event is classified into three different azimuthal regions, relative to the leading track. Assuming $\phi_{\rm trig.}$ as the azimuthal angle for the leading trigger particle and $\phi_{\rm assoc.}$ as the azimuthal angle of the associated particles, the three regions are classified as the following,
\begin{itemize}
\item Near side: $|\phi_{\rm trig.} - \phi_{\rm assoc.}| < \frac{\pi}{3}$ 
\item Away side: $|\phi_{\rm trig.} - \phi_{\rm assoc.}| > \frac{2\pi}{3}$
\item Transverse side: $\frac{\pi}{3} \leq |\phi_{\rm trig.} - \phi_{\rm assoc.}| \leq \frac{2\pi}{3}$
\end{itemize}
The particle production in near side is expected to be dominated by jet fragmentation while the away side region captures some of the back-scattered jets. Both regions are also affected by the underlying event. The transverse region is mostly dominated by the underlying event, with possible contamination from initial- and final-state radiation. This leading-$p_{\rm T}$ selection of  8 $< p_{\rm T} <$ 15 GeV/$c$ ensures that the number density in the transverse region remains nearly independent of leading particle $p_{\rm T}$~\cite{Acharya:2019nqn}. This requirement also ensures that the effects of elliptic flow are negligible, which is important as the goal of the approach is to search for jet quenching effects in small collision systems. The relative transverse activity classifier ($R_{\rm T}$) is defined as,
\begin{eqnarray}
R_{\rm T} = \frac{N_{\rm ch}^{\rm TS}}{\langle N_{\rm ch}^{\rm TS} \rangle},
\label{eq2}
\end{eqnarray}
where $N_{\rm ch}^{\rm TS}$ is the charged-particle multiplicity in the transverse region~\cite{Martin:2016igp,Ortiz:2017jaz}. Using $R_{\rm T}$, one can vary the magnitude of underlying event. For example, as $R_{\rm T} \rightarrow$ 0 are the events are populated with less underlying events and they are expected to be compatible to the jetty events as defined from $S_{\rm 0}^{p_{\rm T} = 1}$.

\section{Results and Discussion}

\subsection{Identified particle production as a function of $S_{\rm 0}^{p_{\rm T}= 1}$}

\begin{figure}[h]
\centering
\includegraphics[width=14pc]{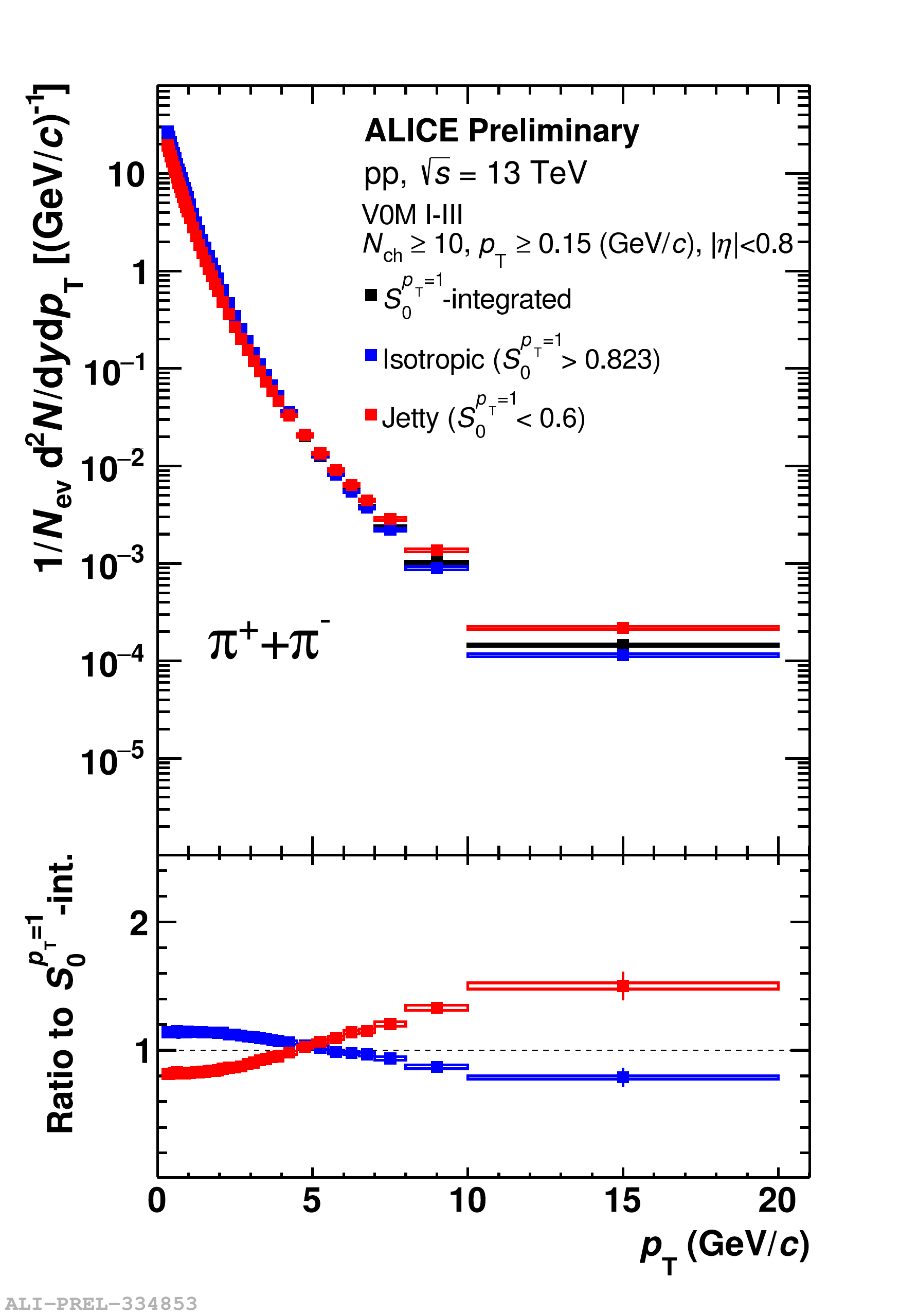}
\includegraphics[width=14pc]{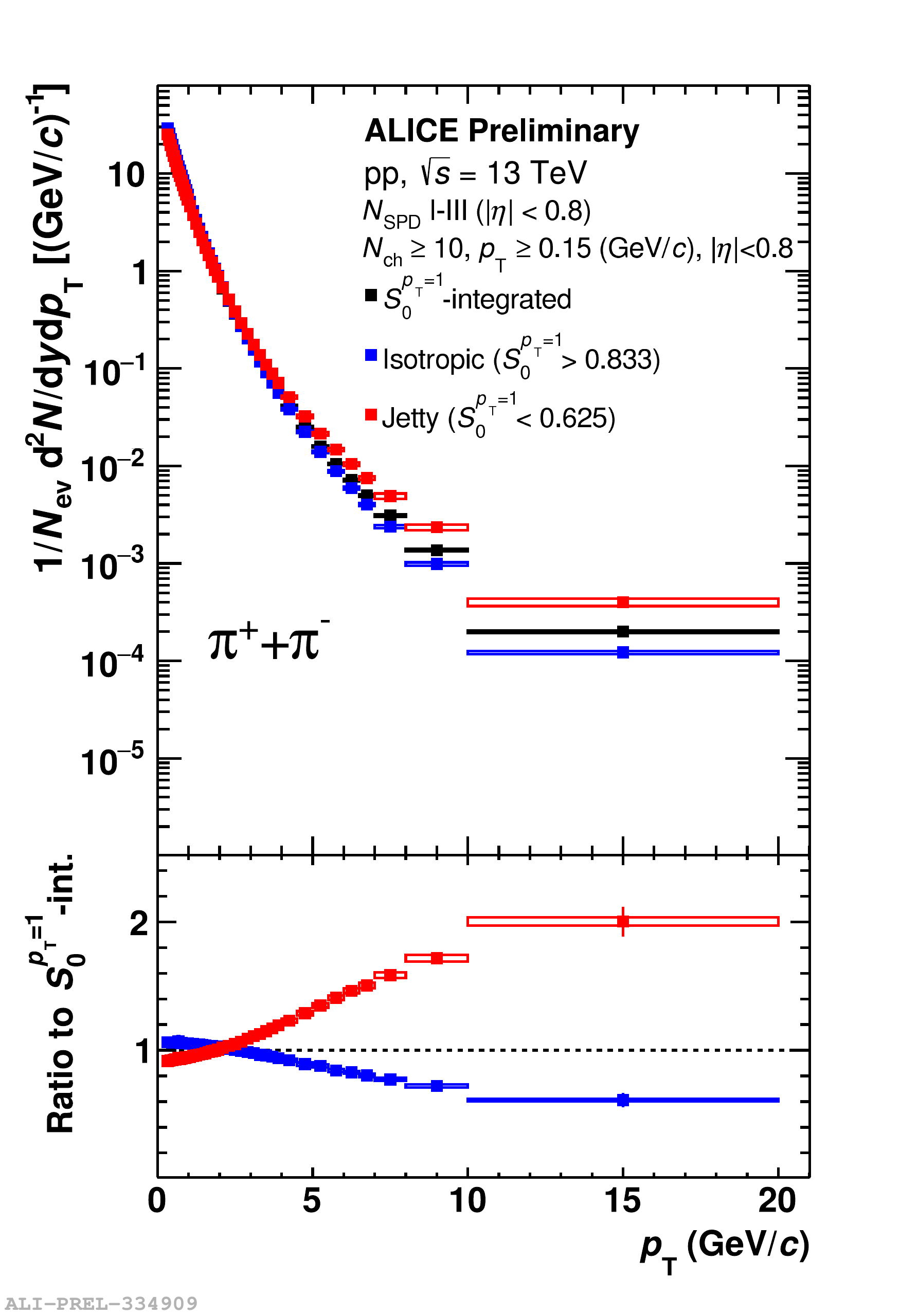}
\caption{\label{fig1} Transverse momentum spectra of pions for different $S_{\rm 0}^{p_{\rm T}= 1}$ classes in pp collisions at $\sqrt{s}$ = 13 TeV. The left (right) panel shows the $p_{\rm T}$ spectra where the multiplicity selection was done using the V0M ($N_{\rm SPD}$) multiplicity estimator. The bottom panels show the ratio of $p_{\rm T}$-spectra from isotropic and jetty events to the $S_{\rm 0}^{p_{\rm T}= 1}$-integrated events.}
\end{figure}

\begin{figure}[h]
\centering
\includegraphics[width=15pc]{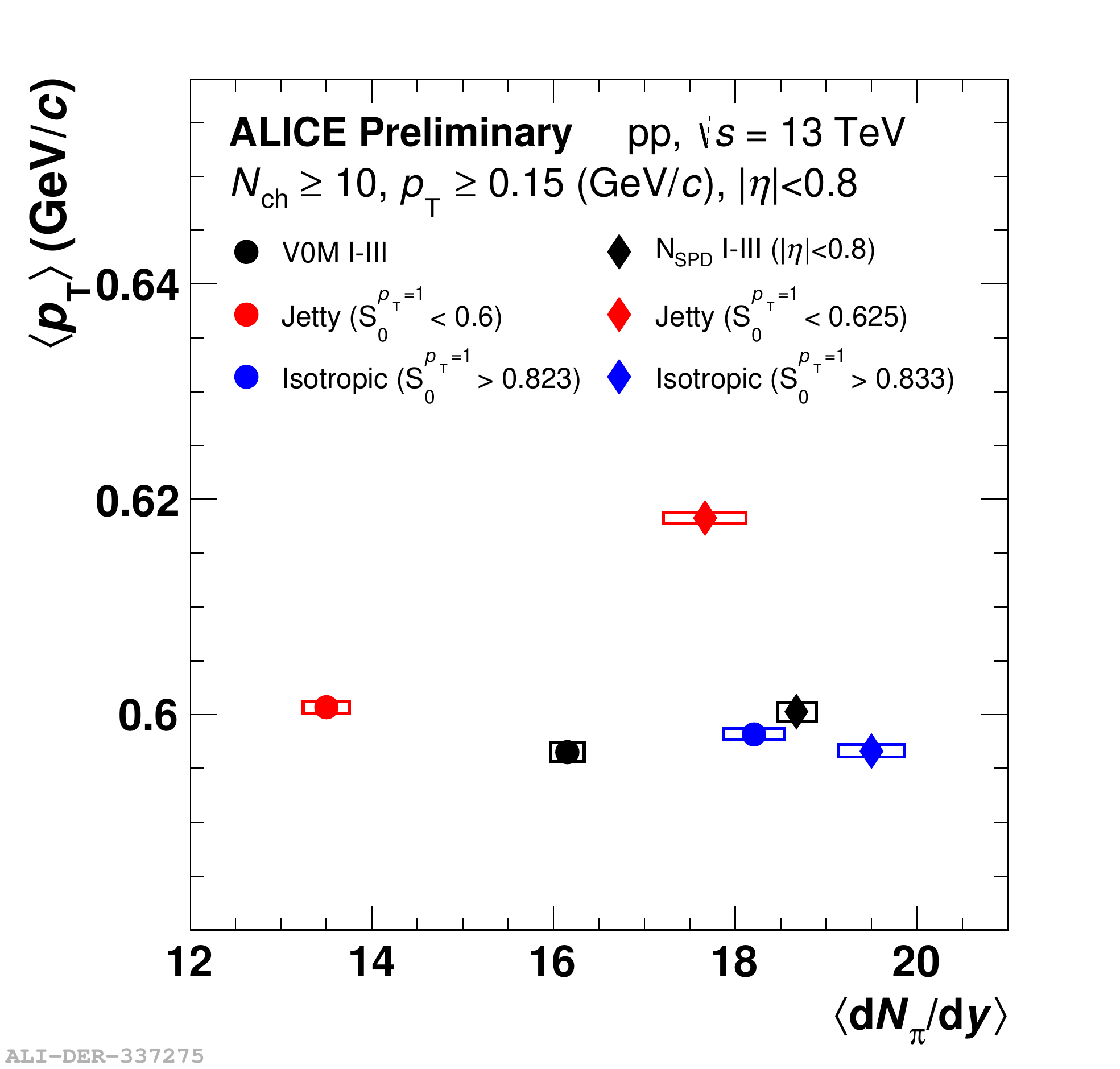}
\begin{minipage}[b]{14pc}\caption{\label{fig2} Pion $\langle p_{\rm T}\rangle$ vs integrated yield for different multiplicity estimators in different $S_{\rm 0}^{p_{\rm T}= 1}$ classes for pp collisions at $\sqrt{s}$ = 13 TeV.}
\end{minipage}
\end{figure}

\begin{figure}[h]
\centering
\includegraphics[width=18pc]{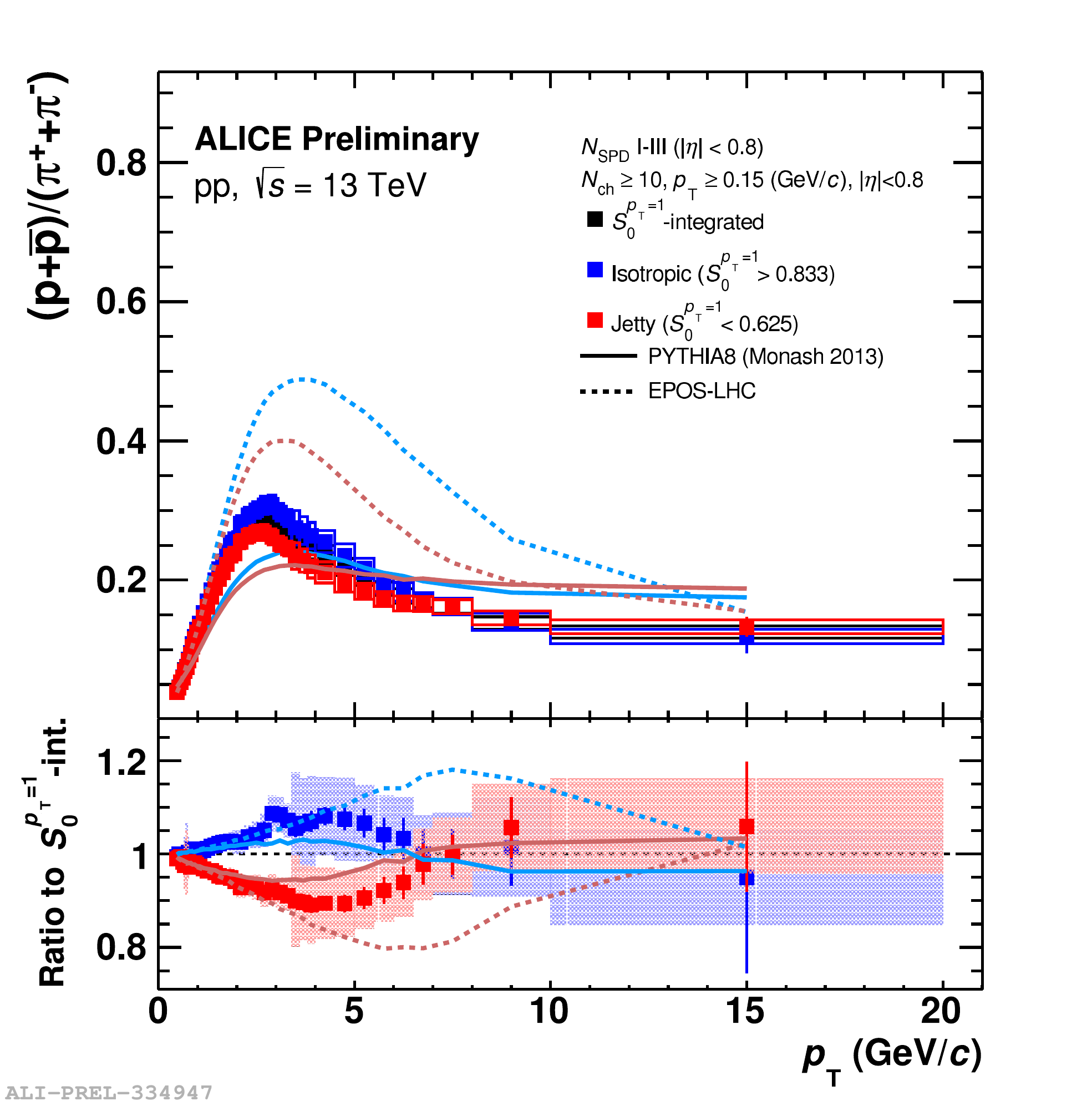}
\includegraphics[width=18pc]{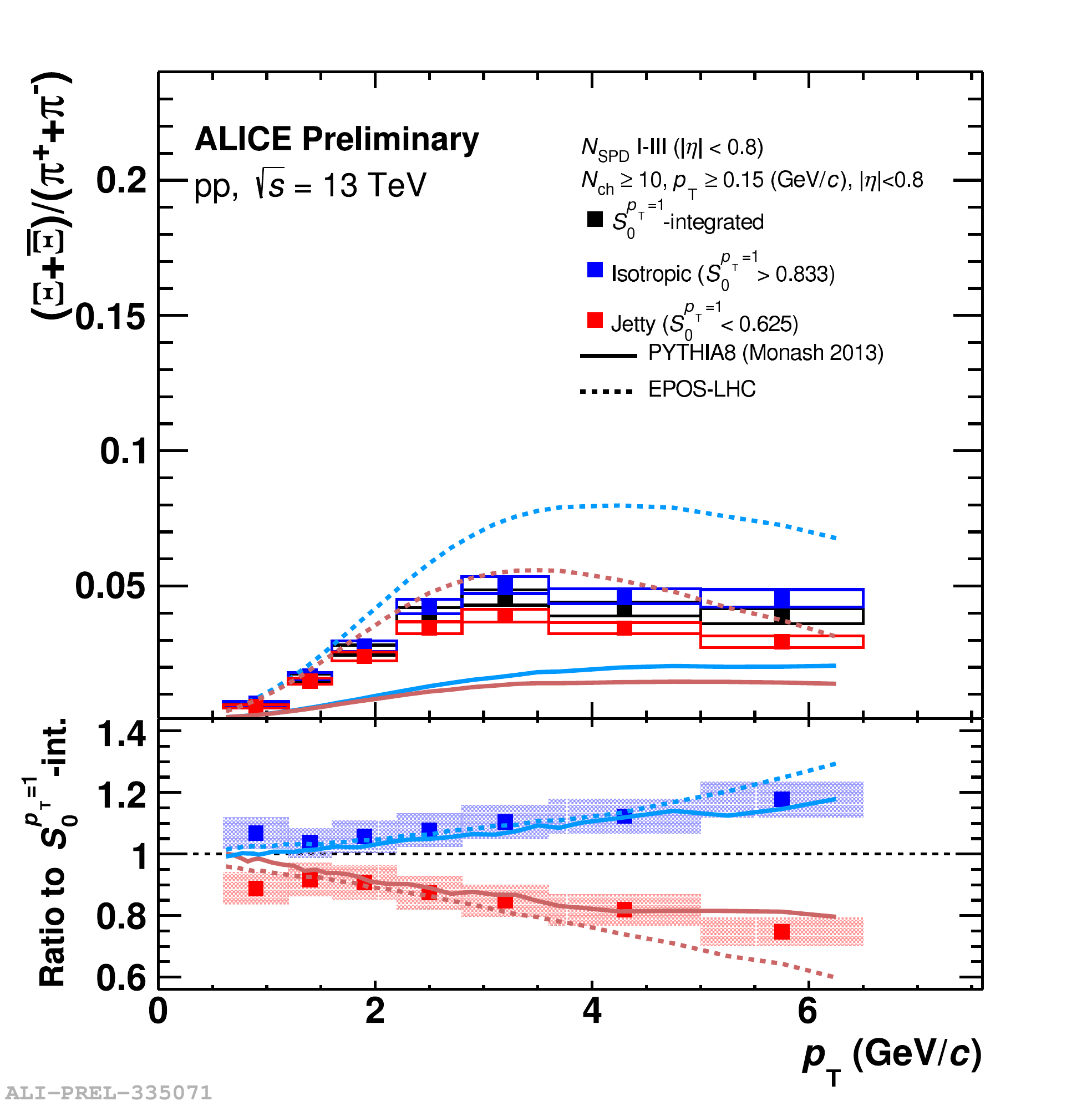}
\caption{\label{fig3} Proton-to-pion (left) and $\Xi$-to-pion (right) ratio as a function $p_{\rm T}$ in different $S_{\rm 0}^{p_{\rm T}= 1}$ classes for pp collisions at $\sqrt{s}$ = 13 TeV where the multiplicity selection was done using the $N_{\rm SPD}$ multiplicity estimator. The bottom panels show the double ratio of particle ratios from isotropic and jetty events to the $S_{\rm 0}^{p_{\rm T}= 1}$-integrated events. The ratios are compared with predictions from PYTHIA 8 and EPOS-LHC.}
\end{figure}

Figure~\ref{fig1} shows the transverse momentum spectra of pions for different $S_{\rm 0}^{p_{\rm T}= 1}$ classes in pp collisions at $\sqrt{s}$ = 13 TeV. The left (right) panel shows the $p_{\rm T}$ spectrum where the multiplicity selection was done using the V0M ($N_{\rm SPD}$) multiplicity estimator. Pions are identified from the specific energy loss ($\rm{d}E/\rm{d}x$) measured by the TPC and time-of-flight (TOF) using the TOF detector. The bottom panels show the ratio of $p_{\rm T}$-spectra from events identified as isotropic and jetty to the $S_{\rm 0}^{p_{\rm T}= 1}$-integrated events. A clear dependence of the $p_{\rm T}$-spectra as a function of $S_{\rm 0}^{p_{\rm T} = 1}$ is observed in the results from both the estimators. However, as can be seen in Fig.~\ref{fig2} (pion $\langle p_{\rm T}\rangle$ vs integrated yield), the V0M-triggered isotropic and jetty events have larger multiplicity difference while the $N_{\rm SPD}$-triggered events disentangle soft and hard events more accurately in small multiplicity gap. 
Figure~\ref{fig3} shows proton-to-pion (left) and $\Xi$-to-pion (right) ratio as a function of $p_{\rm T}$ in different $S_{\rm 0}^{p_{\rm T} = 1}$ classes for pp collisions at $\sqrt{s}$ = 13 TeV, where the multiplicity selection was done using the $N_{\rm SPD}$ multiplicity estimator. The bottom panels show the double ratio of particle ratios from isotropic and jetty events to the $S_{\rm 0}^{p_{\rm T} = 1}$-integrated events. Here, protons are identified using TPC and TOF detectors. The $\Xi$ yields are extracted from invariant mass distributions of its identified decay products after topological selections. The proton-to-pion ratio show a larger enhancement in the intermediate-$p_{\rm T}$ region for isotropic events compared to jetty events, which seems to be reminiscent of a similar effect observed in Pb-Pb collisions~\cite{Acharya:2018orn}. The $\Xi$-to-pion ratio suggests that the strange particle production is higher in isotropic events compared to jetty events. In general, the particle ratios are not described by PYTHIA8 and EPOS-LHC. However, the trend of the double ratios are better described by both PYTHIA8 and EPOS-LHC.

\subsection{System size dependence of charge particle production as a function of $R_{\rm T}$} 
\begin{figure}[h]
\centering
\includegraphics[width=35pc]{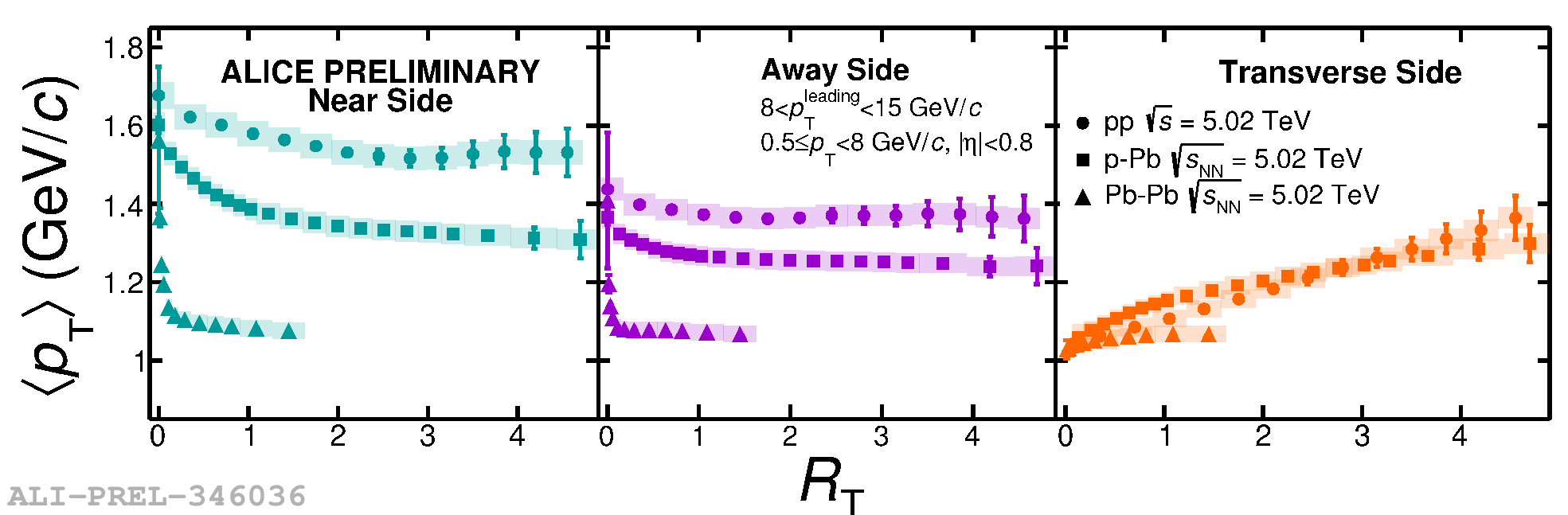}
\caption{\label{fig4} System size dependence of $\langle p_{\rm T}\rangle$ of charged-particles as a function of $R_{\rm T}$ in the near (left), away (middle), and transverse (right) regions.}
\end{figure}

Figure~\ref{fig4} shows the system size dependence of $\langle p_{\rm T}\rangle$ of charged-particles as a function of $R_{\rm T}$ in the near (left), away (middle), and transverse (right) regions. The near and away side $\langle p_{\rm T}\rangle$ for pp and p--Pb collisions decreases at low-$R_{\rm T}$ and saturates for high-$R_{\rm T}$. The $\langle p_{\rm T}\rangle$ for transverse side increases with $R_{\rm T}$. The contribution from the near and away side jet dominates at low-$R_{\rm T}$ and the values are similar for all systems as one would naively expect for $R_{\rm T}\rightarrow$ 0. For large $R_{\rm T}$, the $\langle p_{\rm T}\rangle$ approaches a similar value in all three topological regions for a given system.

\begin{figure}[h]
\centering
\includegraphics[width=16pc]{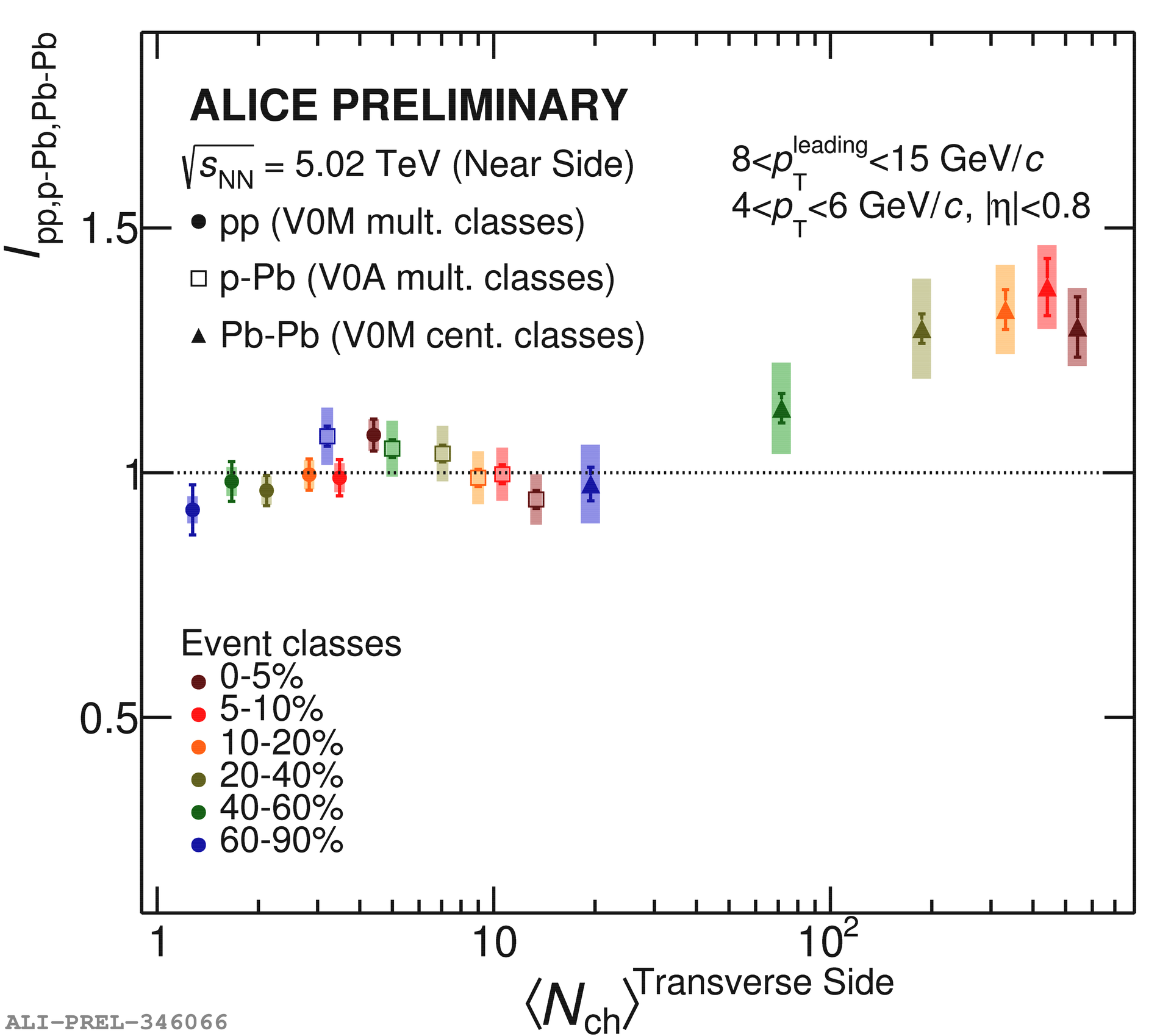}
\includegraphics[width=16pc]{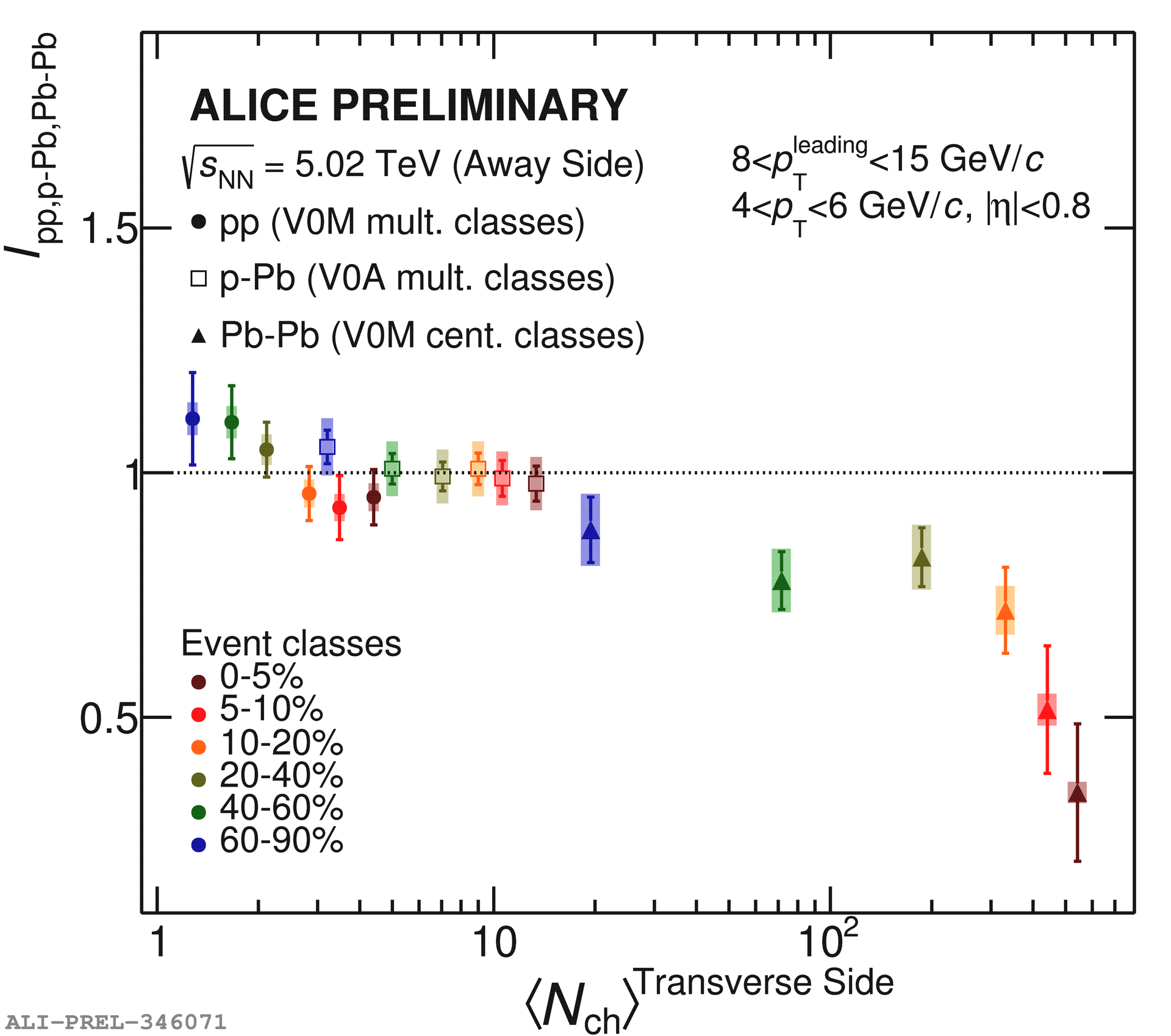}
\caption{\label{fig5} $I_{\rm pp,p-Pb,Pb-Pb}$ as a function of $\langle N_{\rm ch}^{\rm TS} \rangle$ in different V0M/V0A multiplicity classes for the near (left) and away (right) regions in pp, p--Pb, and Pb--Pb collisions at $\sqrt{s_{\rm NN}}$ = 5.02 TeV.}
\end{figure}

To explore the presence of jet-quenching effects we have calculated the $I_{\rm pp,p-Pb,Pb-Pb}$, an observable which is calculated from the yields of different topological regions, as a function of $\langle N_{\rm ch}^{\rm TS} \rangle$ for different V0M/V0A multiplicity classes of pp, p--Pb, and Pb--Pb collisions. The $I_{\rm pp,p-Pb,Pb-Pb}$ is an analogous quantity calculated as in Ref.~\cite{Aamodt:2011vg}. The $I_{\rm pp,p-Pb,Pb-Pb}$ is sensitive to medium effects and any suppression in the away side would indicate the presence of jet quenching, while an enhancement in the near side would indicate the presence of medium effects and a bias due to trigger particle selection.  It is defined as the ratio of yield in the near or away region (after subtraction of underlying events) in different collision systems to the yield in the near or away region in minimum bias pp collisions. It can be expressed as,

\begin{equation}
I_{\rm pp,p-Pb,Pb-Pb} = \frac{Y^{\rm pp,p-Pb,Pb-Pb} - Y^{\rm pp,p-Pb,Pb-Pb}_{\rm TS}}{Y^{\rm pp~min. bias} - Y^{\rm pp~min. bias}_{\rm TS}}.
\end{equation}
Here, $Y$ represents the integrated yield of charged particles. Note that, for these results there is no direct selection on $N_{\rm ch}^{\rm TS}$, as direct selection on $N_{\rm ch}^{\rm TS}$ biases the near and away side yields~\cite{Ortiz:2020dph}. Thus, the events are selected based on the forward rapidity estimator (V0M for pp and Pb--Pb collisions and V0A for p--Pb collisions) and the corresponding $N_{\rm ch}^{\rm TS}$ are calculated. Figure~\ref{fig5} shows the $I_{\rm pp,p-Pb,Pb-Pb}$ for the range 4 $< p_{\rm T}^{assoc.} < $ 6 GeV/$c$ as a function of $\langle N_{\rm ch}^{\rm TS} \rangle$ in different V0M/V0A multiplicity classes for the near (left) and away (right) side in pp, p--Pb, and Pb--Pb collisions at $\sqrt{s_{\rm NN}}$ = 5.02 TeV. The values of $I_{\rm PbPb}$ for central and peripheral Pb-Pb collisions show a similar trend for both the near and away sides as reported by ALICE in Ref.~\cite{Aamodt:2011vg} at Pb--Pb collisions at $\sqrt{s_{\rm NN}}$ = 2.76 TeV. For small collision systems, no enhancement (suppression) of $I_{\rm pp,p-Pb}$ is observed in near (away) sides for pp or p--Pb collisions. This may indicate the absence of jet-quenching effects in small collision systems for the measured  $\langle N_{\rm ch}^{\rm TS} \rangle$ ranges.

\section{Summary}
In summary, using the event shape observables $S_{\rm 0}^{p_{\rm T} = 1}$ and $R_{\rm T}$, the magnitude of the underlying events can be varied to study the events separately with different topological limits (jetty vs isotropic). A clear dependence of light flavor particle $p_{\rm T}$ spectra on $S_{\rm 0}^{p_{\rm T}= 1}$ is observed. The system size study of charged-particle production suggests that the contribution from the near and away side jet dominates at low-$R_{\rm T}$ and the values are similar for all systems, as one would naively expect for $R_{\rm T}\rightarrow$0. For large $R_{\rm T}$, the $\langle p_{\rm T}\rangle$ approaches a similar value in all three topological regions for a given system. In contrast to Pb--Pb collisions, no enhancement (suppression) of $I_{\rm pp,p-Pb}$ is observed in the near (away) sides for pp, p--Pb collisions, which indicates the absence of jet-quenching effects in small collision systems for the measured  $\langle N_{\rm ch}^{\rm TS} \rangle$ ranges.

\section*{Acknowledgments}
S.T. acknowledges the support from CONACyT under the Grant No. A1-S-22917 and postdoctoral fellowship of DGAPA UNAM.

\end{document}